\documentclass[12pt]{article}
\usepackage{amsmath,amssymb,amsthm,amsxtra,overpic,bbm,bm,epsfig,subfigure}
\usepackage{color}
\usepackage{comment}
\usepackage[all,cmtip]{xy}
\usepackage{cite}

\usepackage{slashed,stmaryrd}

\begin{document}

\vspace{0.2cm}

\begin{center}
{\large\bf Constraints on dark matter interactions from the first results of DarkSide-50}
\end{center}

\vspace{0.2cm}

\begin{center}
{\bf Chun-Yuan Li}~$^a$ \footnote{E-mail: lichunyuan@mail.sdu.edu.cn}, \quad {\bf Zong-Guo Si}~$^a$ \footnote{E-mail: zgsi@sdu.edu.cn}, \quad {\bf Yu-Feng Zhou}~$^{b,c}$ \footnote{E-mail: yfzhou@itp.ac.cn}

{$^a$School of Physics, Shandong University, Jinan, Shandong 250100, China}

{$^b$Institute of Theoretical Physics, Chinese Academy of Sciences, Beijing 100190, China}

{$^c$University of Chinese Academy of Sciences, Beijing 100049, China}
\end{center}

\begin{abstract}

In an extended effective operator framework of isospin violating interactions with light mediators, we investigate the compatibility of the candidate signal of the CDMS-II-Si with the latest constraints from DarkSide-50 and XENON-1T, etc. We show that the constraints from DarkSide-50 which utilizes Argon as the target is complementary to that from XENON-1T which utilizes Xenon. Combining the results of the two experiments, we find that for isospin violating interaction with light mediator there is no parameter space which can be compatible with the positive signals from CDMS-II-Si. As a concrete example of this framework, we investigate the dark photon model in detail. We obtain the combined limits on the dark matter mass $m_{\chi}$, the dark photon mass $m_{A'}$, and the kinetic mixing parameter $\varepsilon$ in the dark photon model. The DarkSide-50 gives more stringent upper limits in the region of mediator mass from 0.001 to 1 GeV, for $m_{\chi} \lesssim 6$ GeV in the ($m_{A'}$,$\varepsilon$) plane, and more stringent constraints for $m_{\chi}\lesssim 8$ GeV and $\varepsilon \thicksim 10^{-8}$ in the ($m_{\chi}$,$m_{A'}$) plane.

\end{abstract}

\section{Introduction}

Although the existence of dark matter (DM) has been strongly supported by many astrophysical and cosmological observations, its particle nature remains largely unknown. Weakly interacting massive particles (WIMPs) are the popular candidates of DM~\cite{Jungman:1995df,Bertone:2004pz,Bertone:2016nfn}. In this scenario, DM may have weak interactions with the ordinary matter. At present, numerous underground DM direct detection experiments are built to search for the possible signals arising from the interactions between WIMPs and the Standard Model (SM) particles.

In recent years, several DM direct detection experiments have reported potential signals for WIMPs with masses around few GeV to several ten GeV, including DAMA ~\cite{Bernabei:2008yi,Bernabei:2010mq,Bernabei:2013xsa,Bernabei:2018yyw}, CoGeNT~\cite{Aalseth:2012if,Aalseth:2011wp,Aalseth:2014eft} , and CDMS-II-Si~\cite{cdmsIIsi}, while other experiments only reported upper limits on the scattering cross section. The region of parameters favored by DAMA is excluded by other experiments which with different targets, such as LUX ~\cite{Akerib:2013tjd}, XENON~\cite{xenon1T:2018}, DarkSide ~\cite{DarkSide:2018}, SuperCDMS ~\cite{Agnese:2014aze},and XMASS~\cite{Kobayashi:2018jky}, etc. In order to cross check the potential DM signal of DAMA, several experiments utilize the same NaI(TI) detectors to search for DM, like COSINE~\cite{Ha:2017egw,Ha:2018obm} and ANAIS-112~\cite{Coarasa:2018qzs} etc. Although the ANAIS-112 can detect the annual modulation in the 3$\sigma$ region compatible with the DAMA results, the COSINE-100 still does not observe the event rate that excess over the predicted background. The results of CoGeNT are inconsistent with the negative results from CDEX~\cite{cdex:1404.4946,cdex:1601.04581,cdex:1710.06650,Jiang:2018pic} which utilizes the same type of germanium detector. The CDMS-II-Si reported three WIMP-candidate events. It favors a DM particle mass  $\sim 8.6$ GeV and a spin-independent DM-nucleon scattering cross section $\sim 1.9\times 10^{-41}~\rm{cm}^{2}$. This results are also in tension with the limits of other experiments, when interpreted in terms of DM-nucleus elastic scattering in the simple DM model.

The interpretations of the experimental data involve simplified assumptions. For instance, the interactions between DM particles and target nuclei are often assumed to be isospin conserving, contact, and elastic etc. Simplified assumptions are also adopted on the DM velocity distribution, DM local energy density, nuclear form factors, detector responses, etc. The interpretations of the experimental data can be changed dramatically if some of the assumptions is modified. In order to reconcile the conflicts among the experiments, several mechanisms have been discussed, such as the isospin violating interactions~\cite{Feng:2011-ivdm,Frandsen:2011vidm,Jin:2012jn}, the light WIMPs-nucleus mediators~\cite{S.miao:1412.6220,Geng:2016,Geng:2017}, exothermic scattering ~\cite{Batell:2009vb,Graham:2010ca,Chen:2014tka,Geng:2016,Geng:2017}, the different DM velocity profile~\cite{Mao:2012hf,Lisanti:2010qx,Laha:2016iom,Fowlie:2018svr} and halo-independent~\cite{Purcell:2012sh, lack_darkmatter,Witte:2017qsy}
 ,etc.

In this work, we reinterpret the results from CDMS-II-Si with the new data from DarkSide-50~\cite{DarkSide:2018}, XENON1T~\cite{xenon1T:2018}, CDEX-10~\cite{Jiang:2018pic}, etc., in the extended effective operator framework~\cite{Chang:2009yt,Fan:2010gt,Simone:2013effective,Kumar:2013iva,S.miao:1412.6220} with both isospin violating interactions and light mediators. The effective operator framework is actually the secluded DM scenario, where the DM and the mediator compose a hidden DM sector. The mediator may have sizable coupling with the DM, but its coupling with SM particles is usually very weak by some mechanism. The isospin-violating interaction and light mediators are the popular methods to ameliorate the tensions in the direct detection experiments. In the scenario of isospin violating, the DM particle couples to proton and neutron with different strengths, the possible destructive interference between the two couplings can weaken the bounds from different experiments. The value of $m_{\chi}$ favored by the CDMS-II-Si data increases with the mediator becomes lighter. As a concrete example of this framework, we investigate the dark photon model in detail. The existence of dark sectors is theoretically and phenomenologically motivated, which may contain new particles like dark photon.
In the dark photon model~\cite{Hewett:2012ns,Pospelov:dake_photon,Knapen:dake_photon,Cirelli:dake_photon, Evans:2017kti, Dutra:dake_photon}, we focus on the more stringent constraints on dark photon from DarkSide-50, XENON-1T, etc., in the ($m_{A'}$,$\varepsilon$) plane and the ($m_{\chi}$,$m_{A'}$) plane, respectively.

The paper is organized as follows. In Sec.~2, we present the general framework for DM direct detection. In Sec.~3, we show our results for combining isospin violating interactions and light mediators. In Sec.~4, we use the dark photon model to focus on the data from DarkSide-50 and XENON-1T and give the upper limit in the plane $(m_{A'},\varepsilon)$ and the lower limit in the plane $(m_{\chi}, m_{A'})$. Finally, a short summary is given.

\section{General framework for DM direct detection}

The DM direct detection experiments is one of the most promising techniques to detect particle DM. If the galaxy is filled with WIMPs, many of them should pass through the Earth. As a result, it is  possible to look for the interaction of such particles with ordinary matter. In this paper, we consider the scenario where WIMPs elastically scatter off a target nucleon $N$ in elastic process via exchanging a mediator particle $\phi$ in t-channel. If the mass of mediator $\phi$ is much larger than 3-momentum transfer of the scattering process, the interactions can be effectively described by a set of local Lorentz-invariant operators~ ~\cite{Chang:2009yt,Fan:2010gt,Simone:2013effective,Kumar:2013iva,S.miao:1412.6220}
\begin{equation}
{\cal O}_{i}=\frac{c_{i}}{\Lambda^{2}}(\bar{\chi}\Gamma_{i}\chi)(\bar{N}\Gamma_{i}'N),
\end{equation}
where $c_{i}$ are the coefficients, and $\Lambda$ is the mass scale of the mediator particle. The matrices $\Gamma_{i}$, $\Gamma_{i}'$ are Lorentz-invariant combinations of the Dirac matrices. When the mediator is relatively light, the correction to this effective operator approach can be obtained by a replacement $\Lambda^{2}\rightarrow(q^{2}+m_{\phi}^{2})$, where $q$ is the 3-momentum transfer and $m_{\phi}$ is the mass of the mediator.

The differential cross section for $\chi N$ scattering can be written as
\begin{eqnarray}
  &&\frac{d\sigma_{N}}{dq^{2}}(q^{2},\upsilon)=\frac{\overline{|M_{\chi N}|^{2}}}{64\pi m_{N}^{2}m_{\chi}^{2}\upsilon^{2}},
\end{eqnarray}\\
where ${\overline{|M_{\chi N}|^{2}}}$ is the squared matrix element averaged over the spins of initial particles, and $\upsilon$ is the velocity of the WIMP in the nucleon rest frame. The total DM-nucleon scattering cross section $\sigma_{N}$ are defined by
\begin{eqnarray}
\sigma_{N}(v)=\int_{q_{\rm min}^{2}}^{q_{\rm max}^{2}}dq^{2}\frac{d\sigma_{N}}{dq^{2}}(q^{2},\upsilon),
\end{eqnarray}\\
where $q_{\rm{min}}^{2}$ is an infrared cutoff which value can be related to the energy threshold of DM direct detection experiment, $q_{\rm{max}}^{2}=4\mu^{2}_{\chi N}v^{2}$ is the maximal value allowed by kinematics, and $\mu_{\chi N}$ is the WIMP-nucleon reduced mass.

Since $\sigma_{N}(v)$ is in general a function of $\upsilon$, it is useful to define a velocity-independent cross section $\overline{\sigma}_{N}\equiv \sigma_{N}(v_{\rm ref})$, which is the total cross section at a reference velocity ~$\upsilon_{\rm{ref}}\sim\rm{200~km\cdot{s}^{-1}}$. Thus the differential cross section for $\chi N$ scattering can be rewritten in the conventional form ~\cite{S.miao:1412.6220}

\begin{eqnarray}
\frac{d\sigma_{N}}{dq^{2}}(q^{2},\upsilon)=\frac{\overline{\sigma}_{N}}{4\mu_{\chi N}^{2}\upsilon^{2}}G(q^{2},\upsilon),
\end{eqnarray}
where $G(q^{2},\upsilon)$ is a factor containing the $q^2$-dependence and the rest of $v$-dependence which is defined by
\begin{eqnarray}
G(q^{2},\upsilon)=\frac{(q_{\rm{ref}}^{2}-q_{\rm{min}}^{2})\overline{|M_{\chi N}|^{2}}}{\int_{q_{\rm{min}}^{2}}^{q_{\rm{ref}}^{2}}dq^{2}\overline{|M_{\chi N}(q^{2},\upsilon_{\rm{ref}})|^{2}}},
\end{eqnarray}
and where $q_{\rm{ref}}^{2}\equiv4\mu_{\chi N}^{2}\upsilon_{\rm{ref}}^{2}$.

 Concretely, the corresponding formulae of $G(q^{2},\upsilon)$ can be explicitly obtained for different operator. According to the momentum and velocity dependencies, the effective operators are catalogued into six types. In this paper, we only consider spin-independent scattering, so we only focus on the following three type operators~\cite{S.miao:1412.6220},

{\bf Type-I operators}

\begin{eqnarray}
\emph{O}_{1(1)}&=&\frac{1}{q^{2}+m_{\phi}^{2}}\bar{\chi}\chi\bar{N}N,  ~\emph{O}_{1(2)}=\frac{1}{q^{2}+m_{\phi}^{2}}\bar{\chi}\gamma^{\mu}\chi\bar{N}\gamma_{\mu}N,\nonumber\\
\emph{O}_{1(3)}&=&\frac{2m_{\chi}}{q^{2}+m_{\phi}^{2}}\chi^{\ast}\chi\bar{N}N,
\emph{O}_{1(4)}=\frac{1}{q^{2}+m_{\phi}^{2}}(\chi^{\ast}\overleftrightarrow{\partial_{\mu}}\chi)\bar{N}\gamma^{\mu}N.
\end{eqnarray}

{\bf Type-II operators}

\begin{eqnarray}
\emph{O}_{2(1)}=\frac{1}{q^{2}+m_{\phi}^{2}}\bar{\chi}\gamma^{5}\chi\bar{N}N,
\emph{O}_{2(2)}=\frac{2m_{\chi}}{q^{2}+m_{\phi}^{2}}\chi^{\ast}\chi\bar{N}\gamma^{5}N.
\end{eqnarray}

{\bf Type-III operator}

\begin{eqnarray}
\emph{O}_{3}=\frac{1}{q^{2}+m_{\phi}^{2}}\bar{\chi}\gamma^{\mu}\gamma^{5}\chi\bar{N}\gamma_{\mu} N.
\end{eqnarray}
The factor $G(q^{2},v)$ can be written as~\cite{S.miao:1412.6220,Geng:2017}
\begin{eqnarray}
 G_{1}(q^{2})=\frac{1}{I_{1}(m_{\phi}^{2}+q^{2})^{2}},
  G_{2}(q^{2})=\frac{{q^{2}}/{m_{\phi}^{2}}}{I_{2}(m_{\phi}^{2}+q^{2})^{2}},
   G_{3}(q^{3})=\frac{v_{\bot}^{2}/v_{ref}^{2}}{I_{3}(m_{\phi}^{2}+q^{2})^{2}},
\end{eqnarray}
and
\begin{eqnarray}
 I_{1}=\frac{1}{(1+a)(1+b)},I_{2}=\frac{1}{b-a}\textmd{ln}(\frac{1+b}{1+a})-I_{1},I_{3}=I_{1}-I_{2}/b,
\end{eqnarray}
where $a=q_{\rm{min}}^{2}$, $b=q_{\rm{ref}}^{2}$, $\textit{\textbf{v}}_{\bot}=\textit{\textbf{v}}+\textit{\textbf{q}}/(2\mu_{\chi N})$ is the transverse velocity of the DM particle, and $v^{2}\bot=v^{2}-q^{2}/(4\mu_{\chi N}^{^{2}})$. For type-\rm{III}, the reduced mass $\mu_{\chi N}$ in the expression of $v_{\bot}$ will be replaced by $\mu_{\chi A}$, because the nucleon velocity operator $\emph{\textbf{v}}$ acting on the nucleus wave function will pick up the nucleus mass~\cite{Fitzpatrick:2012ix,S.miao:1412.6220, Geng:2017}.

At nucleus level, for the three type operators, the spin-independent WIMPs-nucleus differential cross section can be written as
\begin{equation}
\frac{d(\sigma_{A})_{i}}{dq^{2}}=\frac{\bar{\sigma}_{p}}{4\mu^{2}_{\chi p}v^{2}}[Z+\xi(A-Z)]^{2}G_{i}(q^{2},v)F^{2}_{A}(q^{2}),
\end{equation}
 where $i=\textrm{I}, \textrm{II}, \textrm{III}$, Z is the number of protons and A is the number of atomic mass number of the target nucleus, $\upsilon$ is the relative velocity of the WIMP in the nuclear rest frame, and $\xi=f_{n}/f_{p}$ where $f_{p}$($f_{n}$) are the DM couplings to protons (neutrons). In the simple model, the scattering is isospin conserving (IC), $\xi\simeq1$. However, in the general model with $\xi\neq1$, the true value of $\sigma_{p}$ (the cross section for the DM particle scattering off a free nucleon) differs from $\sigma_{p}^{IC}$ (the cross section be defined under the assumption that the scattering is isospin conserving) by a factor $K(f_{n}/f_{p})$ which depends on the ratio $f_{n}/f_{p}$ and the target material
  \begin{eqnarray}
 \sigma_{p}=K(f_{n}/f_{p})\sigma_{p}^{IC},
\end{eqnarray}
If $f_{n}/f_{p} < 0$, the interference between the contributions from proton and that from neutron scattering to the value of $K(f_{n}/f_{p})$ is destructive, which can lead to $K(f_{n}/f_{p})\gg 1$. Thus it is possible that the $\sigma_{p}$ value can be a few order of magnitudes larger than $\sigma_{p}^{IC}$. For a given single target material T, the particular value of $f_{n}/f_{p}$ corresponding to the maximal possible value of $K(f_{n}/f_{p})$ can be written by~\cite{Jin:2012jn}
\begin{eqnarray}
 \xi_{T}=-\frac{\sum_{\alpha}\eta_{\alpha}\mu_{\chi m_{A_{\alpha}}}^{2}Z(A_{\alpha}-Z)}{\sum_{\alpha}\eta_{\alpha}\mu_{\chi m_{A_{\alpha}}}^{2}(A_{\alpha}-Z)^{2}},
\end{eqnarray}
where $\mu_{\chi m_{A_{\alpha}}}$ is the reduced mass for the DM and the nucleus with atomic mass number $A_{\alpha}$, $\eta_{\alpha}$ denotes the isotopes abundance, and $\alpha$ denotes different isotopes.  As a concrete example, for Xe and Ar, $\xi_{Xe}\thickapprox$  -0.7 and $\xi_{Ar}\thickapprox$  -0.82. The nuclear form factor $F_{A}^{2}(q^{2})$ is given by~\cite{Lewin:1996Form}
\begin{equation}
F_{A}(q^{2})^{2}=\left(\frac{3j_{1}(qR_{1})}{qR_{1}}\right)^{2}e^{-(qs)^{2}},
\end{equation}
where $j_{1}$ is the first spherical bessel function, $R_{1}=\sqrt[]{R_{A}^{2}-5s^{2}}$ with the effective nuclear radius $R_{A}\simeq1.2A^{1/3}$ fm and $s\simeq$ 1 fm. The factor $G(q^{2})$  reflects the difference between the light mediators interaction and the standard point-like interaction.

The differential recoil event rate per unit detector mass is given by

\begin{eqnarray}
\frac{dR}{dE_{R}}&=&\frac{2N_{T}m_{A}\rho_{\chi}}{m_{\chi}} \int_{v_{\rm{min}}}d^3\emph{\textbf{v}}f(\textbf{\emph{v}})v\frac{d\sigma_{ A}}{dq^{2}},\label{DRER}\\
&=&\frac{\rho_{\chi}\overline{\sigma}_{p}}{2m_{\chi}\mu_{\chi p}^{2}}[Z+\xi(A-Z)]^{2}F^{2}_{A}(E_{R})\int_{\upsilon_{\rm{min}}}G(E_{R},v)\frac{f(\textbf{\textit{v}})}{\upsilon}d^{3}\textbf{\textit{v}},
\end{eqnarray}
where $E_{R}=q^{2}/(2m_{A})$ is the nuclear recoil energy, $m_{A}$ is the mass of the target nucleus, $\rho_{\chi}=\rm{0.3~GeV}\cdot \rm{cm}^{-3}$ is the local WIMPs energy density, $\upsilon_{\rm{min}}=\sqrt{m_{A}E_{R}/(2\mu_{\chi A}^{2})}$ is the minimal velocity that required to generate the recoil energy $E_{R}$ in elastic scattering process, $f(\textbf{\textit{v}})=f_{G}(\textbf{\textit{v}}+\textbf{\textit{v}}_{E};\upsilon_{0},\upsilon_{\rm{esc}})$ is the DM velocity distribution function in the Earth rest frame, and the $f_{G}(\textbf{\textit{v}})$ is the DM velocity distribution in the Galactic halo frame. For the DM velocity profile we adopt the standard halo model~\cite{SHM1209}

\begin{equation}
f_{G}(\textbf{\emph{v}})=\frac{\exp(-v^{2}/v^{2}_{0})}{N_{\rm{esc}}(\pi v_{0}^{2})^{3/2}}\Theta(v_{\rm{esc}}-v),
\end{equation}
where $N_{\rm{esc}}=\textmd{erf}(z)-2z \textmd{exp}(-z^{2})/\pi^{1/2}$
is the normalization constant, with $z\equiv v_{\rm{esc}}/v_{\rm{0}}$,
$v_{0}\approx220~\textmd{km}\cdot \textmd{s}^{-1}$ is the most probable velocity of the DM  particle~\cite{Kerr:1986}, $v_{\textmd{esc}}\approx544~\textmd{km}\cdot \textmd{s}^{-1}$ is Galactic escape velocity from the solar system\cite{Smith:2007}, $\textbf{\textit{v}}_{E}=\textbf{\textit{v}}_{S}+\textbf{\textit{v}}_{ES}\approx\textmd{232}~\textmd{km}\cdot \textmd{s}^{-1}$ is the velocity of the Earth relative to the rest frame of the Galactic halo, $\textbf{\textit{v}}_{S}$ is the velocity of the Sun relative to the rest frame of the Galactic halo, and $\textbf{\textit{v}}_{ES}$ is the velocity of the Earth to the Sun which can lead to annual modulation. The velocity integrals can be read from  ~\cite{S.miao:1412.6220}.

\section{Experimental results and analysis}

In different DM direct detection experiments, the measured signals are different. For instance, the electron-recoil equivalent energy $E_{\rm{ee}}$, the scintillation signal S1, the ionization electron charge signal S2, and the phonon signal, etc. The relation between the measured signals and nuclear recoil energy can be written by
\begin{equation}
s={\rm Q}(E_{R})E_{R}=\nu(E_{R}),
\end{equation}
where $\rm{Q}$ is called quenching factor, for Ge crystal detector, such as CDEX, the $\rm{Q}$ can be read form TRIM software~\cite{Lin:2007ka}. The differential signal event rate can be written by~\cite{S.miao:1412.6220}
\begin{equation}
\frac{dR}{ds}=\int_{0}^{\infty}dE_{R}\epsilon(s)P(s,E_{R})\frac{dR}{dE_{R}},
\end{equation}
where $\epsilon(s)$ is the efficiency of detecting the singles, and $P(s,E_{R})$ is the possibility of probing the single when given the recoil energy $E_{R}$. If a detector with perfect signal resolution, $P(s,E_{R})=\delta(s-\nu(E_{R}))$. The expected number of recoils in the range $[{\rm s_{a},s_{b}}]$ is given by
\begin{eqnarray}
N=\textmd{Ex}\cdot\int_{{\rm s_{a}}}^{{\rm s_{b}}}d{\rm s}\frac{dR}{d{\rm s}},
\end{eqnarray}
where Ex is the exposure given by different experiments. For the target material which composed of multiple elements or isotopes, we sum over the contributions from each component.

We compare the theoretical expected differential signal rate with the energy spectrum or the number of events by experimental measurements, and constrain the parameters related to the WIMPs properties such as $m_{\chi}$, $m_{\phi}$, $\xi$ and $\overline{\sigma}_{p}$ through evaluating the function $\chi^{2}=-\sum2\rm{ln}{\cal L}$, where ${\cal L}$ is likelihood function. If the number of events given in the experiment is relatively small, and the corresponding recoil energy is given, the likelihood function ${\cal L}$ is chosen according to the extended maximum likelihood method~\cite{Geng:2016,Barlow:1990vc}
\begin{eqnarray}
{\cal L}=e^{-(N+B)}\prod_{i}^{n}\left [\left(\frac{dN}{ds}\right)_{i}+\left(\frac{dB}{ds}\right)_{i}\right ]\label{likelihood_1},
\end{eqnarray}
where N and B is the expected total number of signal events from WIMPs and background respectively in the measured range, $(dN/ds)_{i}$ and $(dB/ds)_{i}$ is the differential event rate at the i-th event (i=1,2...n). The $\chi^{2}_{\rm{min}}$ is the minimal value of the $\chi^{2}$, then calculate $\Delta\chi^{2}=\chi^{2}-\chi^{2}_{\rm{min}}$ which is assumed to follow a $\chi^{2}$ distribution. For two degrees of freedom, when $\Delta\chi^{2}=\textmd{4.6 and 6.0}$, the allowed parameter space regions at 90\% and 95\% C.L. For one degrees of freedom, when $\Delta\chi^{2}=\textmd{2.7}$, the allowed parameter space regions at 90\%  C.L.

\subsection{The experimental data}

With the update of many experiments, the largest scale experiments are approaching a background from solar neutrinos that called neutrino wall. At present, the most stringent constraints on the spin-independent cross sections come from the data of DarkSide-50 and XENON-1T. In the $(m_{\chi},\sigma_{p})$ plane, the favored regions from CDMS-II-Si is for few GeV to several ten GeV, so we also focus on this regions. In this work, we shall mainly focus on the interpretation and compatibility of the following experiments.

\begin{itemize}
\item {\bf CDMS-II-Si}. The CDMS-II-Si utilizes silicon detector to measure the ionization electrons signal and the photons signal, at the Soudan Underground Laboratory. The CDMS-II-Si ~\cite{cdmsIIsi} reported an observation of 3 possible DM-induced events with recoil energies at $E_{R}$=8.2, 9.5 and 12.3 keV, respectively, based on a raw exposure of 140.2 kg$\cdot$days. The estimated background from surface event is $0.41^{+0.20}_{-0.08}(\text{stat.})^{+0.28}_{-0.24}(\text{syst.})$. Other known backgrounds from neutrons and $^{206}\text{Pb}$ are $<0.13$ and $<0.08$ at the $90\%$ C.L., respectively. We adopt the acceptance efficiency from Fig.~1 of Ref~\cite{cdmsIIsi}, and assume the resolution to be perfect. We use the extended maximum  likelihood function (\ref {likelihood_1}).

\item {\bf XENON-1T}. The XENON-1T utilizes a liquid xenon time projection chamber with an exposure of 1.3 $\times$ 278.8 t$\cdot$ days, at the Gran Sasso underground laboratory in Italy. This DM search combines data from two science runs, SR0 ~\cite{xenon1T:1705.06655} and SR1. The event found in ~\cite{xenon1T:1705.06655} did not pass event selection criteria in later analysis.
The total efficiencies are shown in Fig.1 of Ref.~\cite{xenon1T:2018}. The data of DM search in the fiducial mass are shown in Fig.3 of Ref.~\cite{xenon1T:2018}. Table I of Ref.\cite{xenon1T:2018} shows the number of events predicted in these regions by the post-fit models as well as the number of observed events after unblinding. The differential signal event rate in dual-phase xenon experiments can be written by~\cite{Geng:2016,xenon:1103}
\begin{equation}
\frac{dR}{d{\rm S1}}=\sum_{0}^{\infty}\varepsilon({\rm S1})\textmd{Gauss}({\rm S1}|n,\sqrt{n}\sigma_{\textmd{PMT}})\int_{0}^{\infty}\textmd{Poiss}(n|\nu(E_{R}))
\varepsilon_{{\rm S2}}(E_{R})\frac{dR}{dE_{R}}dE_{R}\label{DSER},
\end{equation}
where S1 is the primary scintillation light, S2 is the ionization charge, $\varepsilon({\rm S1})$ is the S1 detection efficiency, $\varepsilon_{{\rm S2}}(E_{R})$ is an efficiency cutoff, n is the PE number, $\nu_{E_{R}}$ is the expected number of PE for a given recoil energy $E_{R}$. The corresponding single photoelectron resolution is between (35-40)\% ~\cite{XENON1T:2015,xenon1T:1512.07501}.

\item {\bf DarkSide-50}. The DarkSide-50 utilizes dual-phase argon time projection chamber to search DM at Laboratorio Nazionale del Gran Sasso in Italy. The detection mechanism is similar to that of the liquid xenon experiment. The bulk of the background for the DarkSide-50 experiment is from ordinary radioactivity, producing ionizing electron recoils. This can be identified and rejected by looking at the shape of the S1 signal of each event. Previous Dark Matter searches with DarkSide use pulse shape discrimination on the primary scintillation signals S1 to suppress electron recoil backgrounds. Those analyses were sensitive to the DM masses above a few tens of GeV. The DarkSide-50~\cite{DarkSide:2018} presents a search for Dark Matter with a much lower recoil analysis threshold. Their analysis is sensitive to DM masses down to 1.8 GeV. From the analysis of the last 500 days of exposure, the DarkSide-50 $N_{e^{-}}$ spectra at low recoil energy can be read from Fig.7 of Ref.~\cite{DarkSide:2018}.

\item {\bf CDEX-10 and PandaX-$\textrm{II}$}. CDEX and PandaX are two direct detector experiments of China, both located at the China Jinping Underground Laboratory. The CDEX-10~\cite{Jiang:2018pic} utilizes a P-type point-contact germanium detector with an exposure of 102.8 kg$\cdot$ days and the analysis threshold of 160 eVee. The lower reach of $m_{\chi}$ is extended to 2 GeV, and the date can be read from the Fig.3 of Ref.~\cite{Jiang:2018pic}. The PandaX-II~\cite{pandax2:run9} utilizes dual-phase xenon time-projection chamber, with the exposure of 2.6$\times$104 kg$\cdot$ days. One event was found below the nucleon recoil median curve with an expected background event number of $2.4^{+0.7}_{-0.7}$, in the S1 range 3 to 45 PE. The detection efficiency from the black solid curve and the dashed line at 1.1 keVnr indicates the cutoff used in the WIMP limit setting in Fig.2 of Ref.~\cite{pandax2:run9}. We read  $\nu(E_{R})$ from Fig.4 of Ref.~\cite{pandax2:run9} by digitizing (${\rm S1}, E_{R}$) values along the (red) centroid NR curve. The expected spectrum of PandaX-II is also calculated using the function of Eq.~\eqref {DSER}.

\end{itemize}

\subsection{Results}
In this section, we consider several combinations of the two typical mechanisms that isospin-violation and light mediator, in order to make the CDMS-II-Si data be compatible with the other null experiments. In our analysis, the isospin violation parameter is fixed at $\xi$ = -0.7 or -0.82, and the mediator mass is fixed at $m_{\phi}=\textrm{200 or 1 MeV}$. We choose $q_{\rm {min}}$ to be zero for simplicity. In the following, we investigate the 68\% and 90\% C.L. favored regions from CDMS-II-Si \cite{cdmsIIsi} as well as the 90\% C.L. upper limits from XENON-1T~\cite{xenon1T:2018}, DarkSide-50~\cite{DarkSide:2018}, CDEX-10~\cite{Jiang:2018pic}, and PandaX-$\textrm{II}$~\cite{pandax2:run9}, in the $(m_{\chi}, \sigma_{p})$ plane.

\begin{figure}[!htbp]
\centering
\subfigure[]{
\label{fig:subfig:a}
\includegraphics[height=5cm,width=5cm]{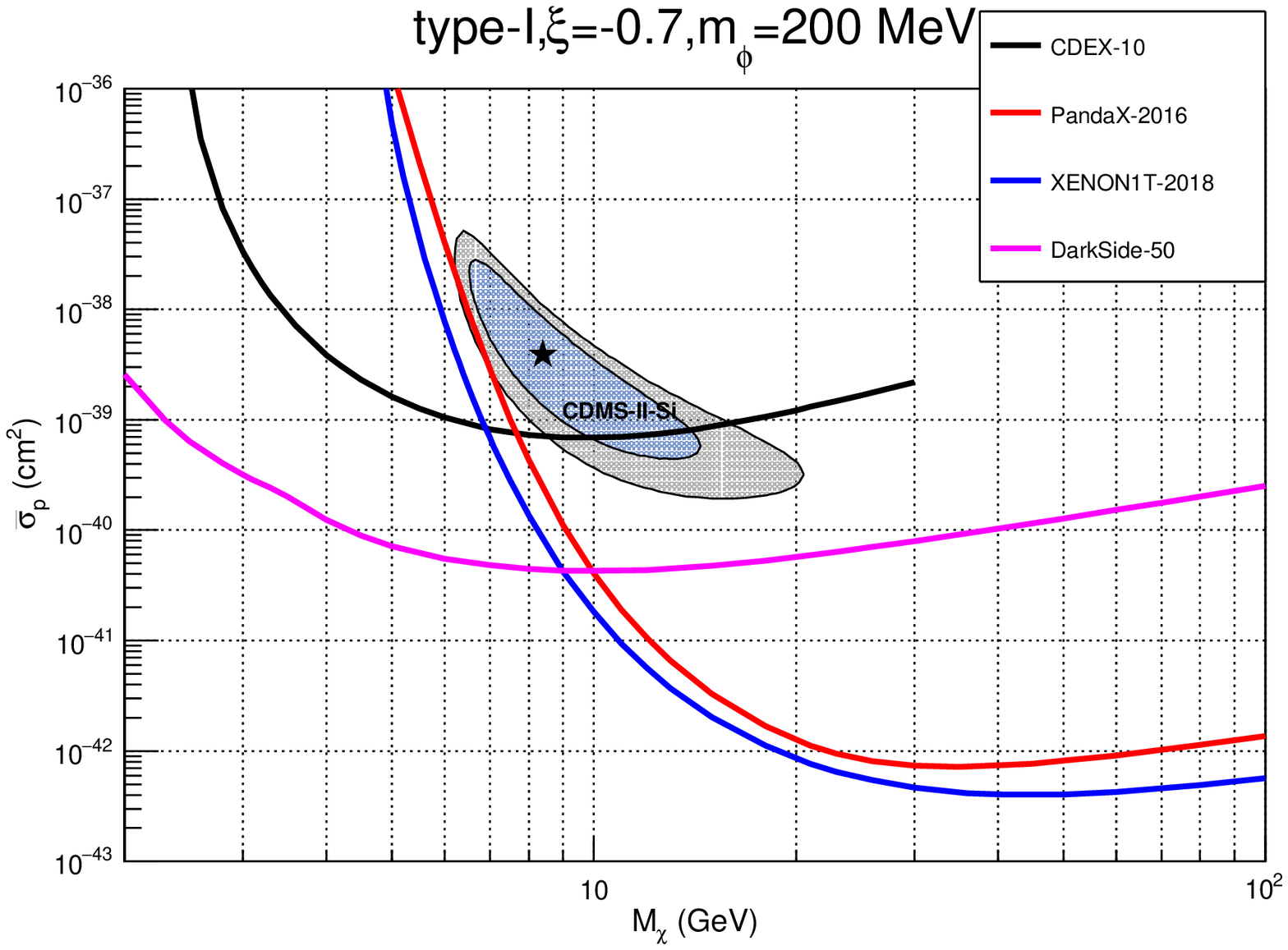}}
\hspace{0.1pt}
 \subfigure[]{
\label{fig:subfig:b}
\includegraphics[height=5cm,width=5cm]{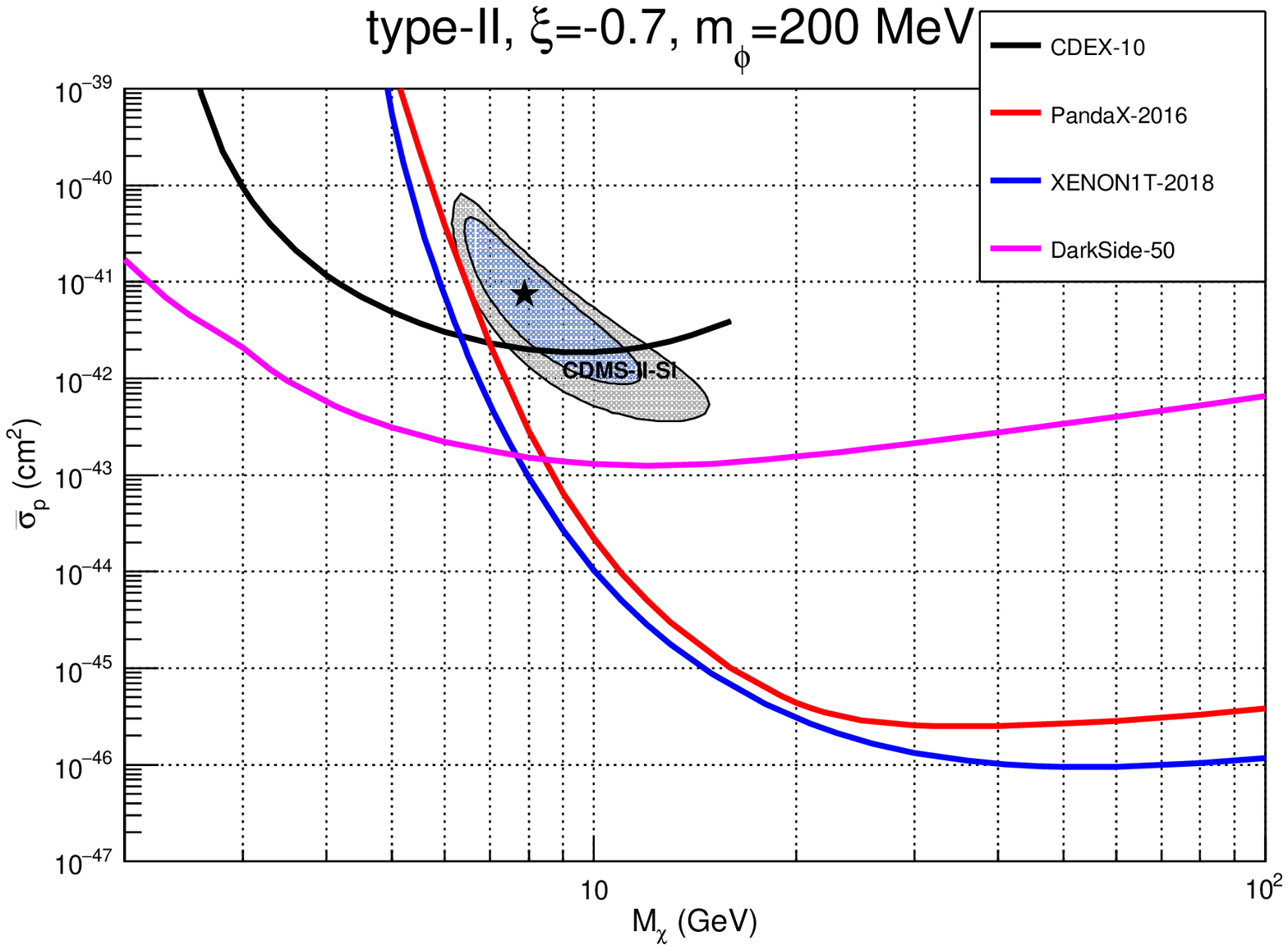}}
\hspace{0.1pt}
\subfigure[]{
\label{fig:subfig:c}
\includegraphics[height=5cm,width=5cm]{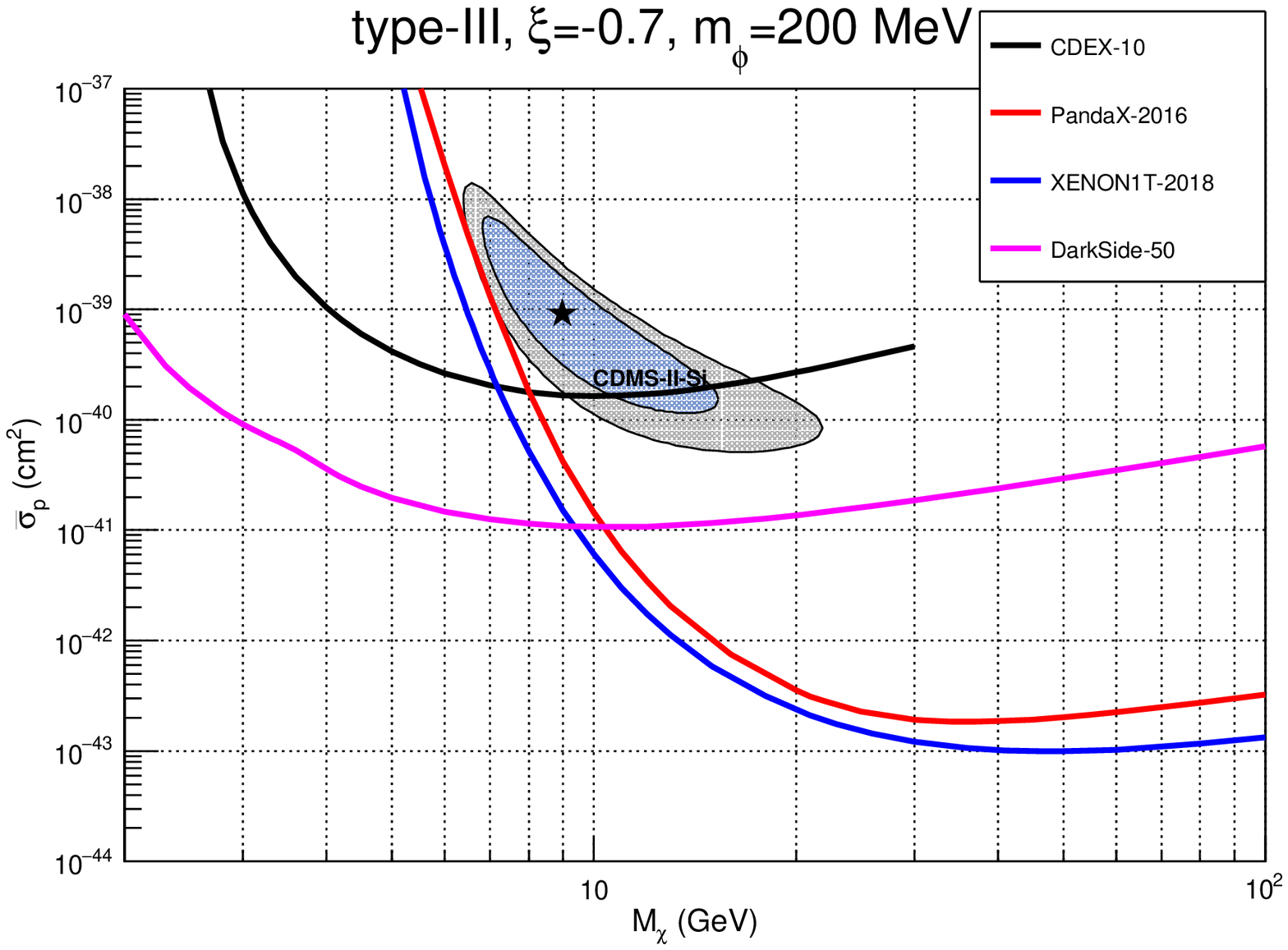}}
\caption{The 68\% and 90\% C.L. favored regions from CMDS-II-Si ~\cite{cdmsIIsi}, as well as 90\% C.L. upper limits from XENON-1T~\cite{xenon1T:2018}, DarkSide-50~\cite{DarkSide:2018}, CDEX-10~\cite{Jiang:2018pic} and  PandaX-$\textrm{II}$~\cite{pandax2:run9} in the $(m_{\chi},\sigma_{p})$ plane. For type-I, II, III operators(from left to right) with $m_{\phi}=\textrm{200 MeV}$, and the isospin violation parameter is fixed at $\xi$ = -0.7.}
\label{fig:1}
\end{figure}

\begin{figure}[!htbp]
\centering
\subfigure[]{
\label{fig:subfig:d}
\includegraphics[height=5cm,width=5cm]{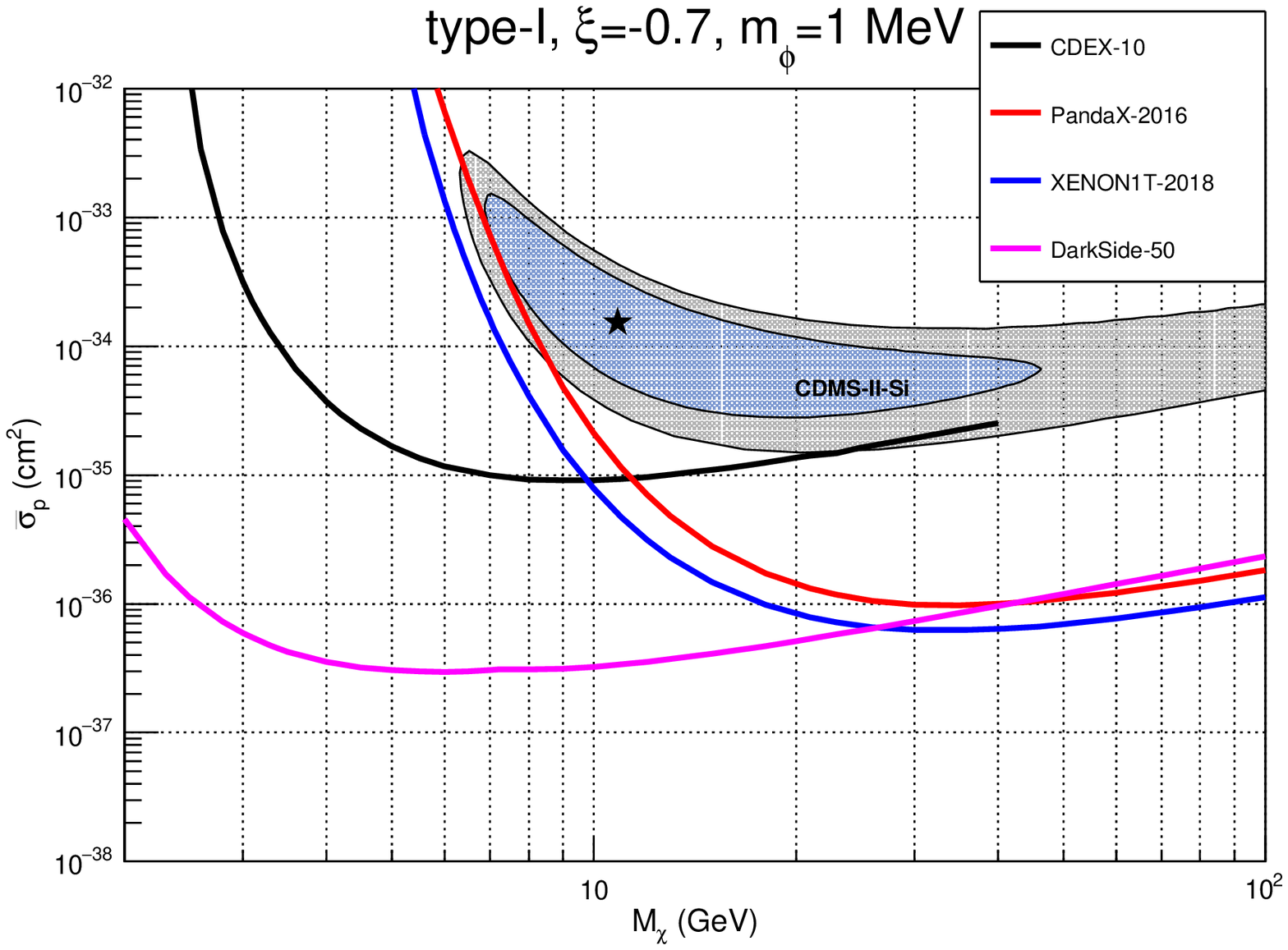}}
\hspace{0.1pt}
 \subfigure[]{
\label{fig:subfig:e}
\includegraphics[height=5cm,width=5cm]{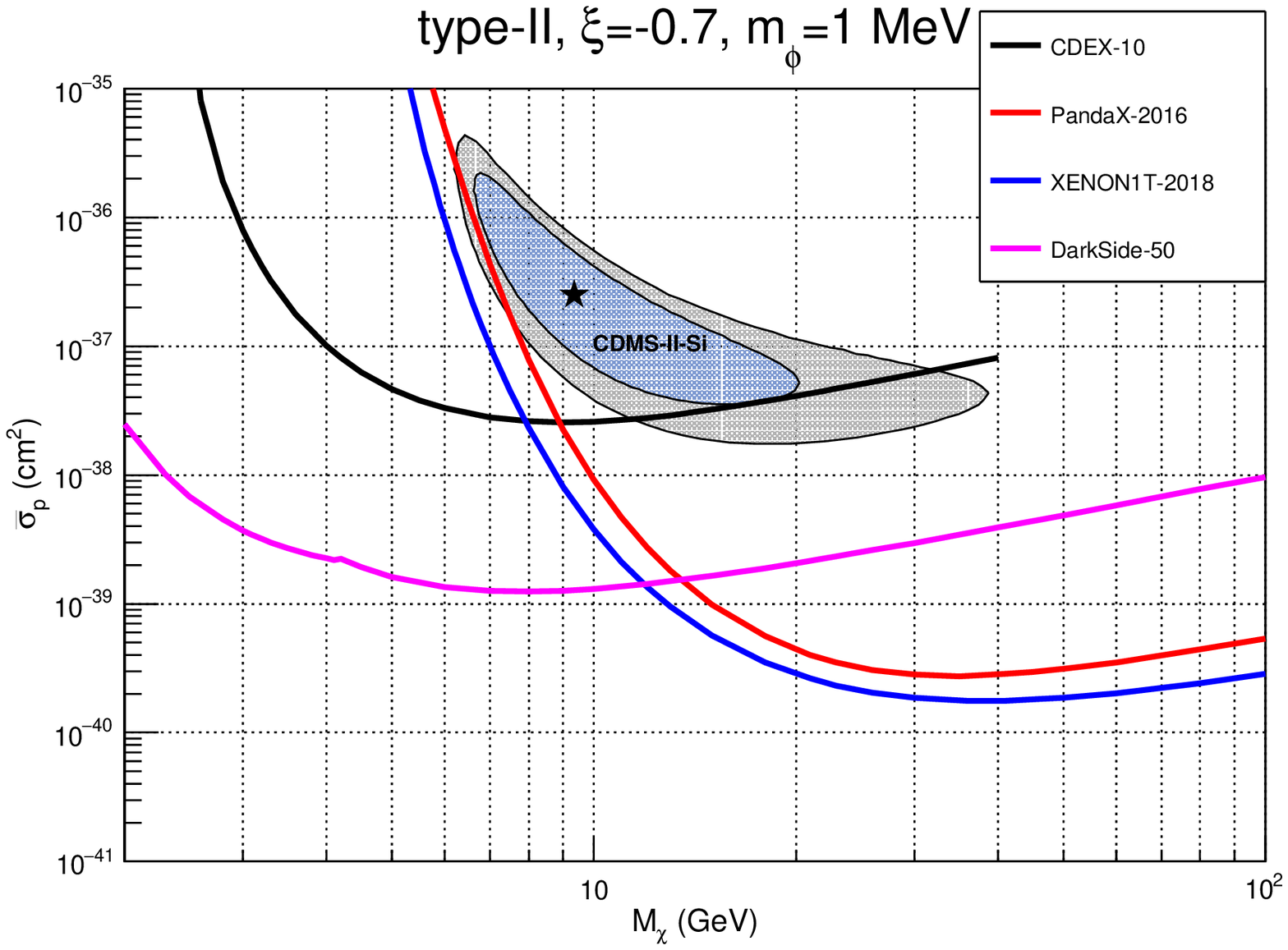}}
\hspace{0.1pt}
\subfigure[]{
\label{fig:subfig:f}
\includegraphics[height=5cm,width=5cm]{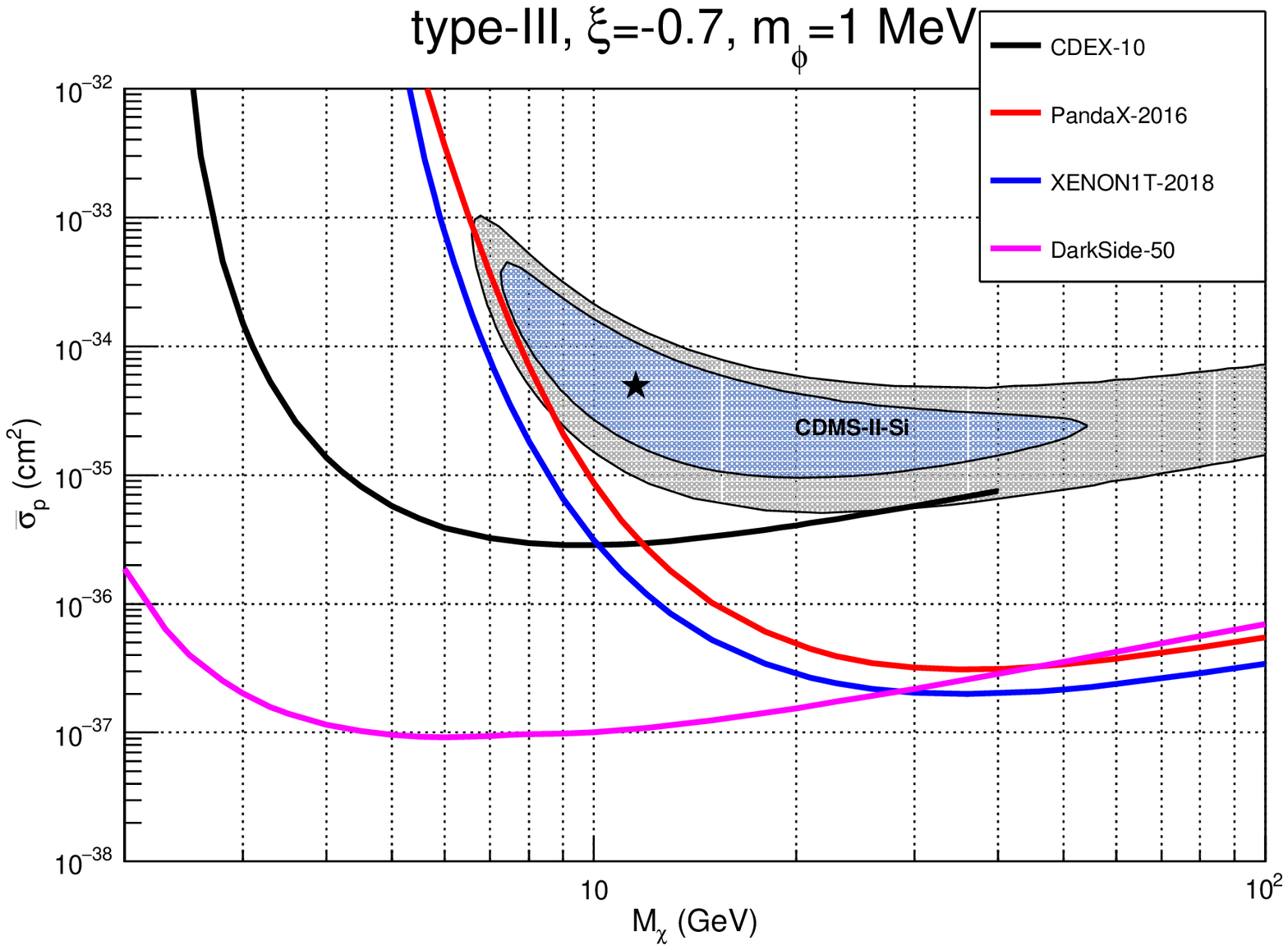}}
\caption{ Legend is the same as Fig. 1 but for $m_{\phi}$ = 1 MeV and $\xi$ = -0.7.}
\label{fig:2}
\end{figure}

\begin{figure}[!htbp]
\centering
\subfigure[]{
\label{fig:subfig:h}
\includegraphics[height=5cm,width=5cm]{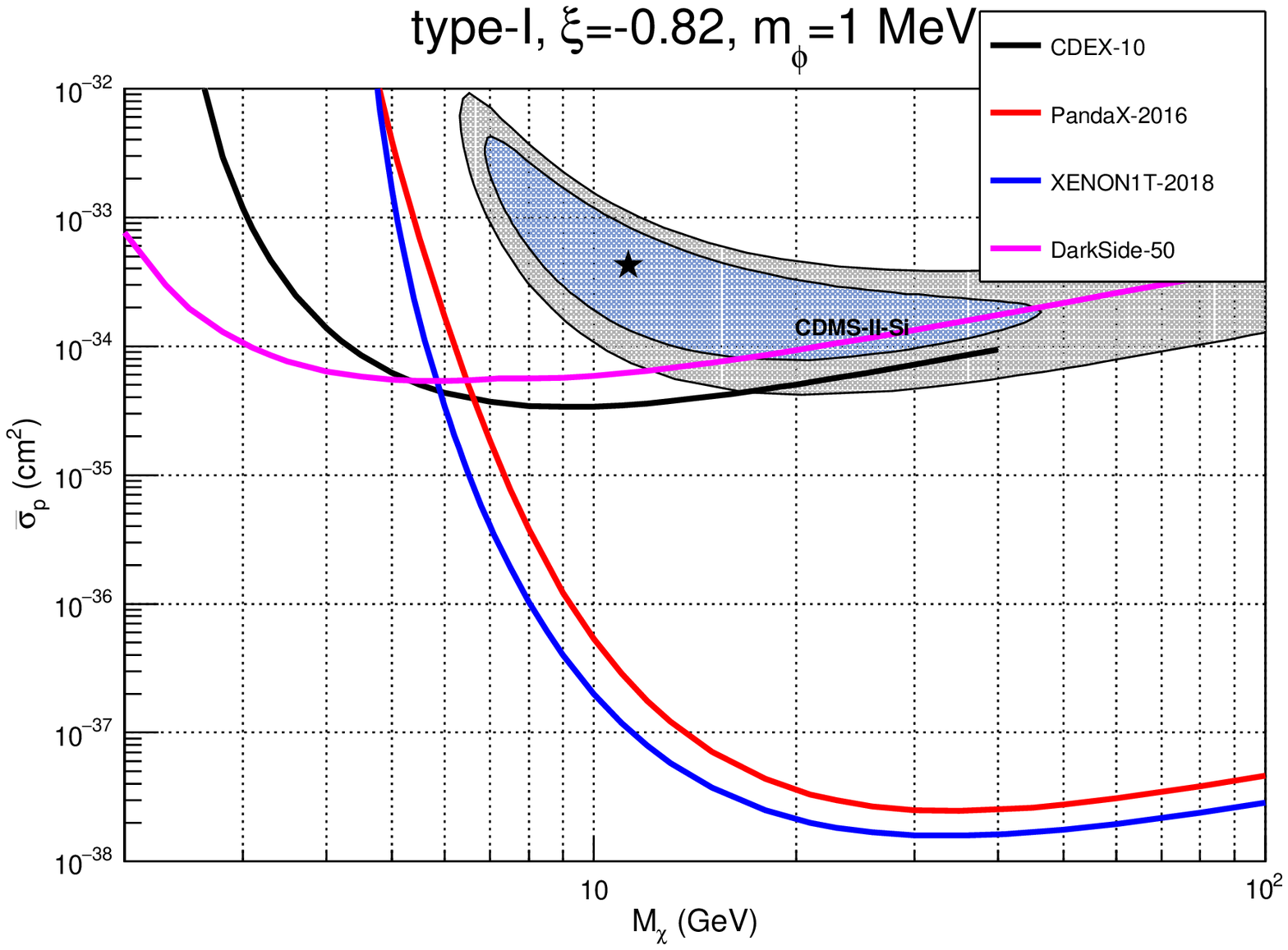}}
\hspace{0.1pt}
 \subfigure[]{
\label{fig:subfig:i}
\includegraphics[height=5cm,width=5cm]{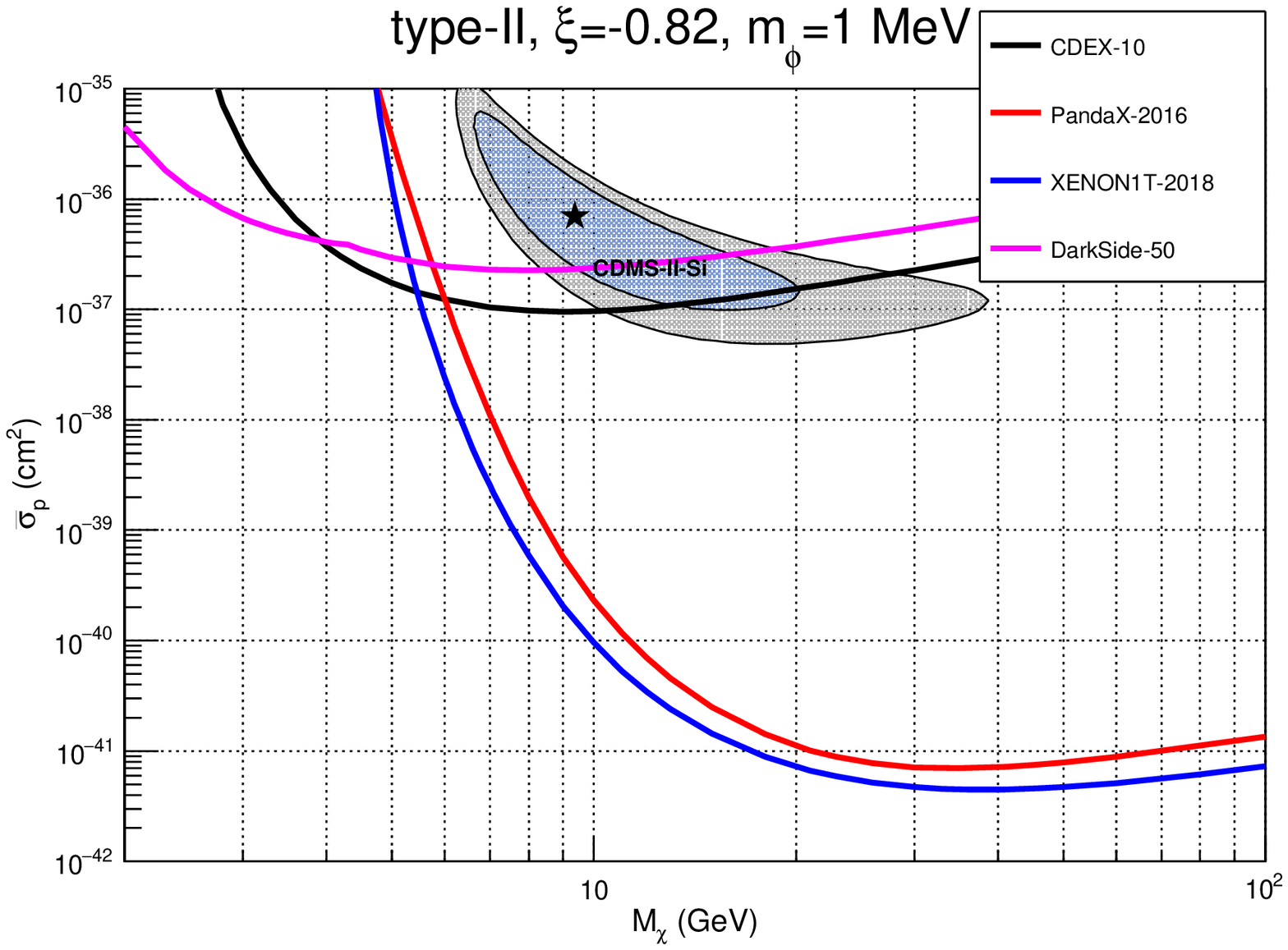}}
\hspace{0.1pt}
\subfigure[]{
\label{fig:subfig:j}
\includegraphics[height=5cm,width=5cm]{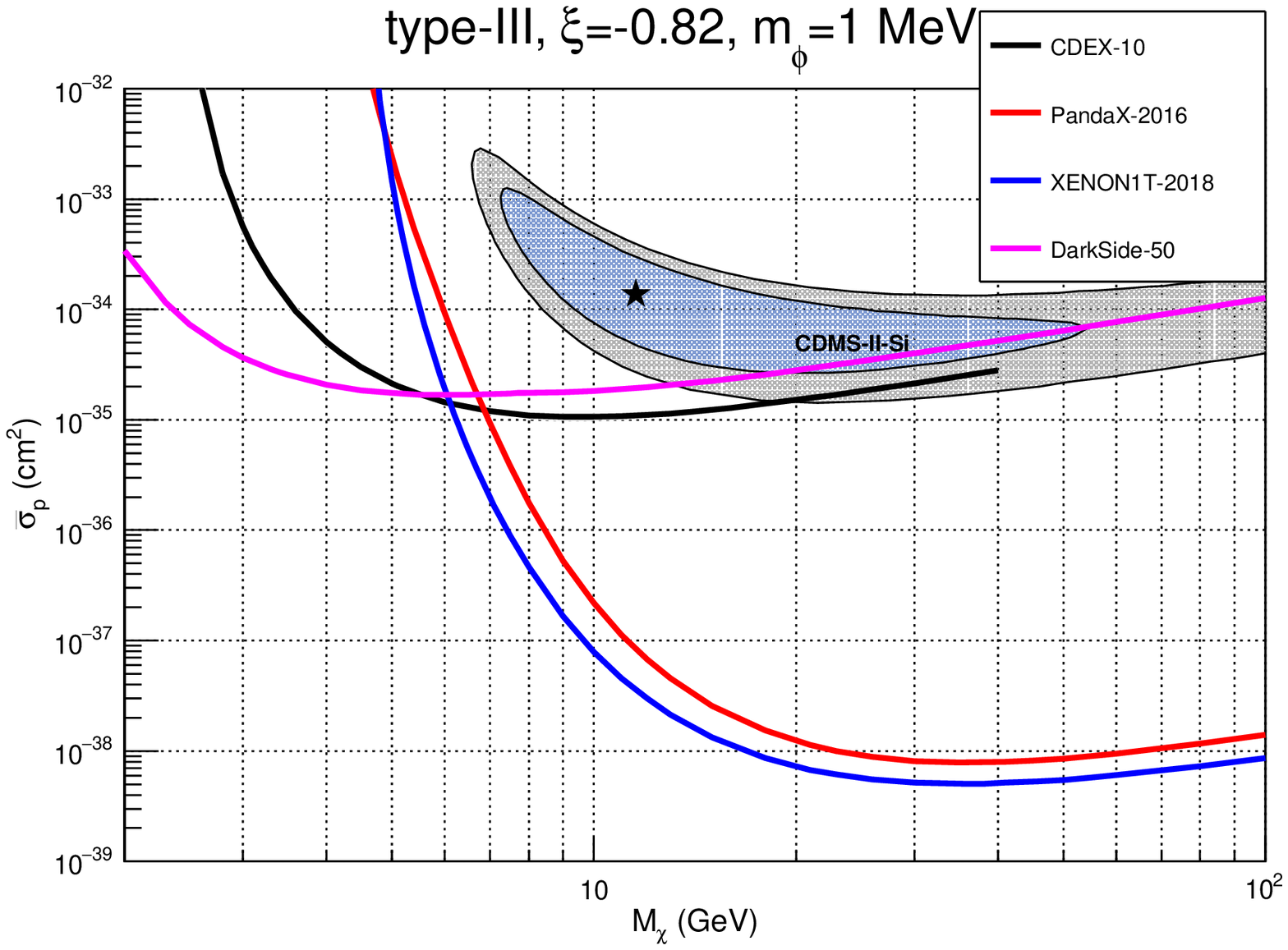}}
\caption{ Legend is the same as Fig. 1 but for  $m_{\phi}$ = 1 MeV and $\xi$ = -0.82.}
\label{fig:3}
\end{figure}

 \begin{itemize}
\item \textbf{} We extract the favored regions with $\xi$ = -0.7 and $m_{\phi}$ = 200 MeV from the experiments mentioned above. The corresponding results are displayed in  Fig.~\ref{fig:1}. For type-I operator, although the XENON-1T constraint is maximally weakened for the case $\xi$ = -0.7, the 68\% and 90\% C.L. favored regions from CDMS-II-Si is excluded by XENON-1T. For the region $m_{x}\lesssim$ 9 GeV, the DarkSide-50 can give the most stringent constraints. For type-II and type-III operators, the conclusions are similar. There are about three order of magnitude difference between the value of the favored region and the upper limits of type-I (type-III) and that of type-II.

\item\textbf{} In order to investigate the effect of light mediator to relax the tension between these experimental results, we fix $m_{\phi}$ = 1 MeV and $\xi$ = -0.7, and display the results in Fig.~\ref{fig:2}. It is clear that for the three operators the value of $m_{\chi}$ favored by the CDMS-II-Si data increases when the mediator becomes lighter, and the upper limits from XENON-1T and PandaX-II become weaker towards high DM particle mass. XENON-1T has a slightly weaker limit in low-mass range, while DarkSide-50 can give more strict limits. For the type-I operator with $m_{\phi}$ = 1 MeV, DarkSide-50 can give the most stringent constraints in the region $m_{\chi}\lesssim$ 20 GeV. The favored regions from CDMS-II-Si are excluded by DarkSide-50 and XENON-1T.

\item\textbf{} The results with the $\xi$ = -0.82 and $m_{\phi}$ = 1 MeV are shown in Fig.~\ref{fig:3}. The constraint of DarkSide-50 is maximally weakened, and gives the most stringent constraints in lower mass region. The restrictions on CDMS-II-Si become weaker in the high mass range. For the type-I operators with $m_{\phi}$ = 1 MeV, DarkSide-50 can give most strict limits in the range $m_{\chi}\lesssim$ 5.5 GeV, but cannot exclude the favored regions for $m_{\chi}\gtrsim$ 11 GeV. For the other two operators, one can obtain the similar conclusions. Compared with  Fig.~\ref{fig:2}, the constraint from XENON-1T becomes more stringent. The favored regions from CDMS-II-Si are still excluded by DarkSide-50 and XENON-1T.

\end{itemize}

 Focusing on the complementary constraints from XENON-1T and DarkSide-50, it is found that in the standard halo model the isospin violation cannot make CDMS-II-Si be consistent with all the other experiments anymore.

\newpage
\section{Dark Photon Model}

In the previous section, we investigate the 68\% and 90\% C.L. favored regions from CMDS-II-Si ~\cite{cdmsIIsi}, as well as 90\% C.L. upper limits from XENON-1T~\cite{xenon1T:2018}, DarkSide-50~\cite{DarkSide:2018}, CDEX-10~\cite{Jiang:2018pic} and PandaX-$\textrm{II}$~\cite{pandax2:run9} in the general DM model with light mediator. As a concrete example, the light mediator may be a dark photon.  In this work, we illustrate the power of the latest data in constraining the dark photon model which is well-motivated and has been extensively studied. To introduce an extra $U(1)^{\prime}$ gauge group is an simple extension of the Standard Model. The dark photon $A'$ arises from the the extra $U(1)^{\prime}$ gauge group, and can mix with the ordinary photon via a kinetic mixing terms~\cite{Pospelov:dake_photon,Knapen:dake_photon,Cirelli:dake_photon ,Evans:2017kti, Dutra:dake_photon}. After the kinetic mixing terms diagonalization, the Lagrangian of the dark photon model is given by~\cite{Alexander:dake_photon,Dutra:dake_photon}

\begin{eqnarray}
{\cal L}&\supset& \sum_{i}\bar{f_{i}}(-eq_{f_{i}}\gamma^{\mu}A_{\mu}-\varepsilon eq_{f_{i}}\gamma^{\mu}A'_{\mu}-m_{f_{i}})f_{i}+\bar{\chi}(-g_{\chi}\gamma^{\mu}A'_{\mu}-m_{DM})\chi\nonumber\\
&&-\frac{1}{4}F_{\mu\nu}F^{\mu\nu}-\frac{1}{4}F^{\prime}_{\mu\nu}F^{\prime\mu\nu}+\frac{1}{2}m_{A'}^{2}A'^{2},
\end{eqnarray}\\
where $m_{f_{i}}$, $m_{\chi}$ and $m_{A'}$ denote the masses of the SM fermion, DM particle and the dark photon, respectively. $F^{\mu\nu}$ and $F^{\prime\mu\nu}$ are the fields strength of the ordinary photon $A$ and that of the dark photon $A'$, $\varepsilon$ is the kinetic mixing parameter in the physical basis, $g_{\chi}$ is the coupling between the dark photon and the dark sector, $\alpha_{\chi}=g_{\chi}^{2}/(4\pi)$ is the dark fine structure constant.

In the dark photon model, the differential cross section for $\chi N$ scattering at the non-relativistic limit can be written as~\cite{Cirelli:dake_photon,dmphoton_detection,dmphoton_detection1}
\begin{eqnarray}
 \frac{d\sigma}{dE_{R}}(v,E_{R})=\frac{8\pi\alpha_{em}\alpha_{\chi}\epsilon^{2}m_{T}}{(2m_{T}E_{R}+m_{A^{'}}^{2})^{2}}\frac{1}{v^{}}Z_{T}^{2}F^{2}(2m_{T}E_{R}),
\end{eqnarray}
where $m_{T}$ is the mass of the target nucleus, $Z_{T}$ is the number of protons in the target nuclei, $F(2m_{T}E_{R})$ is the Helm form factor~\cite{Lewin:1996Form,Helm:form}, and $\alpha_{em}=e^{2}/4\pi$ is the electromagnetic fine structure constant.  The dark fine structure constant $\alpha_{\chi}$ can be determined by the relic abundance of DM.
We take the DM particle to be a Dirac fermion, and consider the case that the present DM abundance is set by thermal freeze out related to the annihilation process $\chi \overline{\chi} \rightarrow A^{'} A^{'}$.  The cross section of DM annihilation can be written as ~\cite{Liu:2014cma}

 \begin{eqnarray}
\langle\sigma \upsilon\rangle\approx\frac{\pi \alpha_{\chi}^{2}}{m_{\chi}^{2}}\frac{(1-m_{A'}^{2}/m_{\chi}^{2})^{3/2}}{[1-m_{A'}^{2}/(2m_{\chi}^{2})]^{2}}.
\end{eqnarray}
Reproducing the observed DM relic abundance of $\Omega_{\chi}h^{2}\approx 0.11$ requires $\langle\sigma \upsilon\rangle \approx $ 2.2 ${\rm cm^{3}/s}$~\cite{Steigman:2012nb}. In the limit of $m_{\chi}\gg m_{A^{'}}$, one finds
 \begin{eqnarray}
\alpha_{\chi}^{{\rm F}} \approx 0.0245\left(\frac{m_{\chi}}{{\rm TeV}}\right).
\end{eqnarray}

The most stringent bounds on $\alpha_{\chi}$ can also come from the imprint of DM annihilation products on the cosmic microwave background (CMB)~\cite{Adams:1998nr,Padmanabhan:2005es,Slatyer:2015jla}. For this aim, the DM abundance is set by non-thermal dynamics and allow $\alpha_{\chi}$ to take its maximal experimentally-allowed value. The corresponding maximum coupling $\alpha_{\chi}$ can be read from ~\cite{Feng:2016ijc}
\begin{eqnarray}
\alpha_{\chi}^{{\rm CMB}}\lesssim 0.17\left(\frac{m_{\chi}}{{\rm TeV}}\right)^{1.61}.
\end{eqnarray}

\begin{figure}[!htbp]
\centering
\subfigure[]{
\label{a}
\includegraphics[height=8cm,width=8cm]{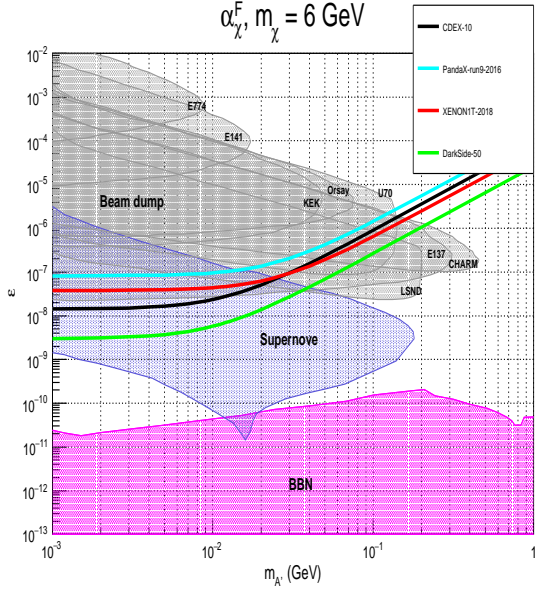}}
\hspace{0.1pt}
 \subfigure[]{
\label{b}
\includegraphics[height=8cm,width=8cm]{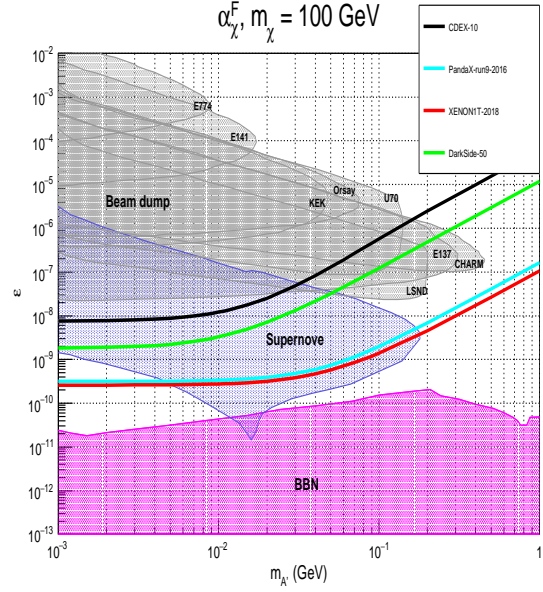}}
\subfigure[]{
\label{c}
\includegraphics[height=8cm,width=8cm]{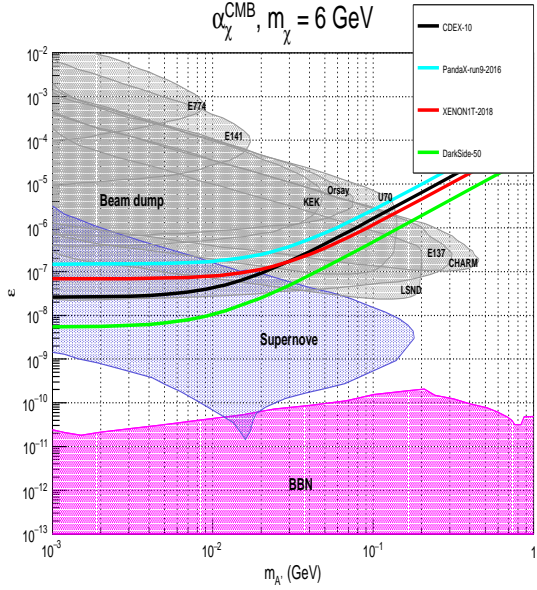}}
\hspace{0.1pt}
 \subfigure[]{
\label{d}
\includegraphics[height=8cm,width=8cm]{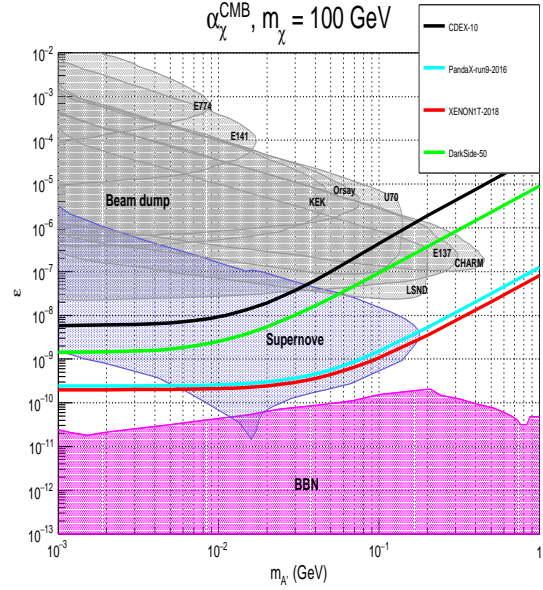}}
\caption{Constraints on the kinetic mixing.The excluded regions in the plane($m_{A'}, \varepsilon$), taking into account several beam dump experiments (gray shaded areas), supernove (blue shaded areas), BBN arguments (pink shaded areas), and the 90\% C.L.  upper limits in the ($m_{A'}$, $\varepsilon$) plane from XENON-1T~\cite{xenon1T:2018}, DarkSide-50~\cite{DarkSide:2018}, CDEX-10~\cite{cdex:1710.06650} and PandaX-$\textrm{II}$~\cite{pandax2:run9} for two DM masses: 6 GeV and 100 GeV, with different DM fine structure constant $\alpha_{\chi}^{{\rm F}}$(upper), $\alpha_{\chi}^{{\rm CMB}}$(down).}
\label{darkphoton1} 
\end{figure}

Before discussing the constraints from direct detection experiments in the dark photon model, we briefly overview the constraints from other experiments.
\begin{itemize}
\item \textbf{Beam dump experiments.}
In electron beam dump experiments, the dark photons can be emitted in a process which is similar to ordinary bremsstrahlung due to the kinetic mixing. The detector is placed behind a sufficiently long shield to suppress the SM background. Dark photons can traverse this shielding due to their weak interactions with the SM particles and can then be detected through their decay into leptons~\cite{Andreas:2012mt,Essig:2013lka}. Several photon beam dump experiments were operated in the last decades, such as experiments E141 ~\cite{Riordan:1987aw} and E137 ~\cite{Bjorken:1988as} at SLAC, the E774~ \cite{Bross:1989mp} experiment at Fermilab, an experiment at KEK ~\cite{Konaka:1986cb} and an experiment in Orsay ~\cite{Davier:1989wz}. Proton beam dump experiments can also be used to search for dark photons which decay through visible channels, the exclusion area from the reinterpretation of LSND ~\cite{ Batell:2009di,Essig:2010gu} at LANSCE, $\nu$-Cal I~\cite{Blumlein:1990ay,Blumlein:2011mv} at the U70 accelerator at IHEP Serpukhov, and CHARM ~\cite{Bergsma:1985is,Gninenko:2012eq} at CERN are also show in Fig.~\ref{darkphoton1}.

\item\textbf{Supernova Bounds}. Light dark photons with a mixing parameter in the range $10^{-10}<\varepsilon<10^{-6}$ are constrained by the neutrino energy spectrum observed after the explosion of supernova SN1987A~\cite{Hirata:1987hu,Bionta:1987qt,Dent:2012mx,Kazanas:2014mca}. In the standard picture, the vast majority of energy that liberated from the collapsing star leaves the supernova in the form of neutrinos. If dark photon are produced in large numbers, it can provides a new cooling mechanism. The cooling of the supernova core becomes more efficient if enough SM photons from the explosion oscillate into $A'$, and if enough $A'$ escape the supernova without further interacting nor decaying~\cite{Raffelt:1996wa,Cirelli:2016rnw}. We show the fiducial exclusion from ~\cite{Chang:2016ntp} as a blue shaded region in Fig.~\ref{darkphoton1}.

\item\textbf{Cosmology}. In the past two decades, there has been impressive progress in our understanding of the cosmological history of the universe. The kinetic mixing portal is one of the few renormalizable interaction channels between the SM and a neutral hidden sector. We can make new constraints on the parameter of dark photon, by calculating the abundance of these dark photons in the early universe and exploring the impact of late decays on BBN and the CMB. We also show the disfavoured BBN area ~\cite{Cirelli:2016rnw} as a pink shaded region in Fig.~\ref{darkphoton1}.

\end{itemize}

We explore the constraints on the parameter space of dark photon from the experiments of DM direct detection, on the case of spin independent DM-nucleus scattering. Fig.~\ref{darkphoton1} shows the excluded regions in the ($m_{A'}$,$\varepsilon$) plane. The solid lines are the constraints on the kinetic mixing from the experiments of DM direct detection. The shaded areas are the excluded regions from beam dump experiments, supernova and BBN arguments etc. The dark fine structure constant $\alpha_{\chi}$ is determined by the abundance of DM. The results of the constraints with $\alpha_{\chi}=\alpha_{\chi}^{{\rm F}}~ (\alpha_{\chi}^{{\rm CMB}})$ are displayed in the upper (down) panel of Fig.~\ref{darkphoton1}. Our analysis show that the mixing parameters $\varepsilon$ is allowed to be around $10^{-10}$ with the mediator mass range from 0.001 to 1 GeV. When we fix the DM mass at 6 (100) GeV, the upper limit with $\alpha_{\chi}^{{\rm F}}$ ($\alpha_{\chi}^{{\rm CMB}}$) is more stricter than the upper limit with $\alpha_{\chi}^{{\rm CMB}}$ ($\alpha_{\chi}^{{\rm F}}$). The upper limit obtained from different experiments have different sensitivities for various DM mass. The upper limits of the PandaX-II and XENON-1T are more sensitive to the DM mass, while the DarkSide-50 can give more stringent upper limits for $m_{\chi} \lesssim 6$ GeV and 0.001 GeV $< m_{A'} <$ 1 GeV.

\begin{figure}[!htbp]
\centering
\subfigure[]{
\label{1}
\includegraphics[height=8cm,width=8cm]{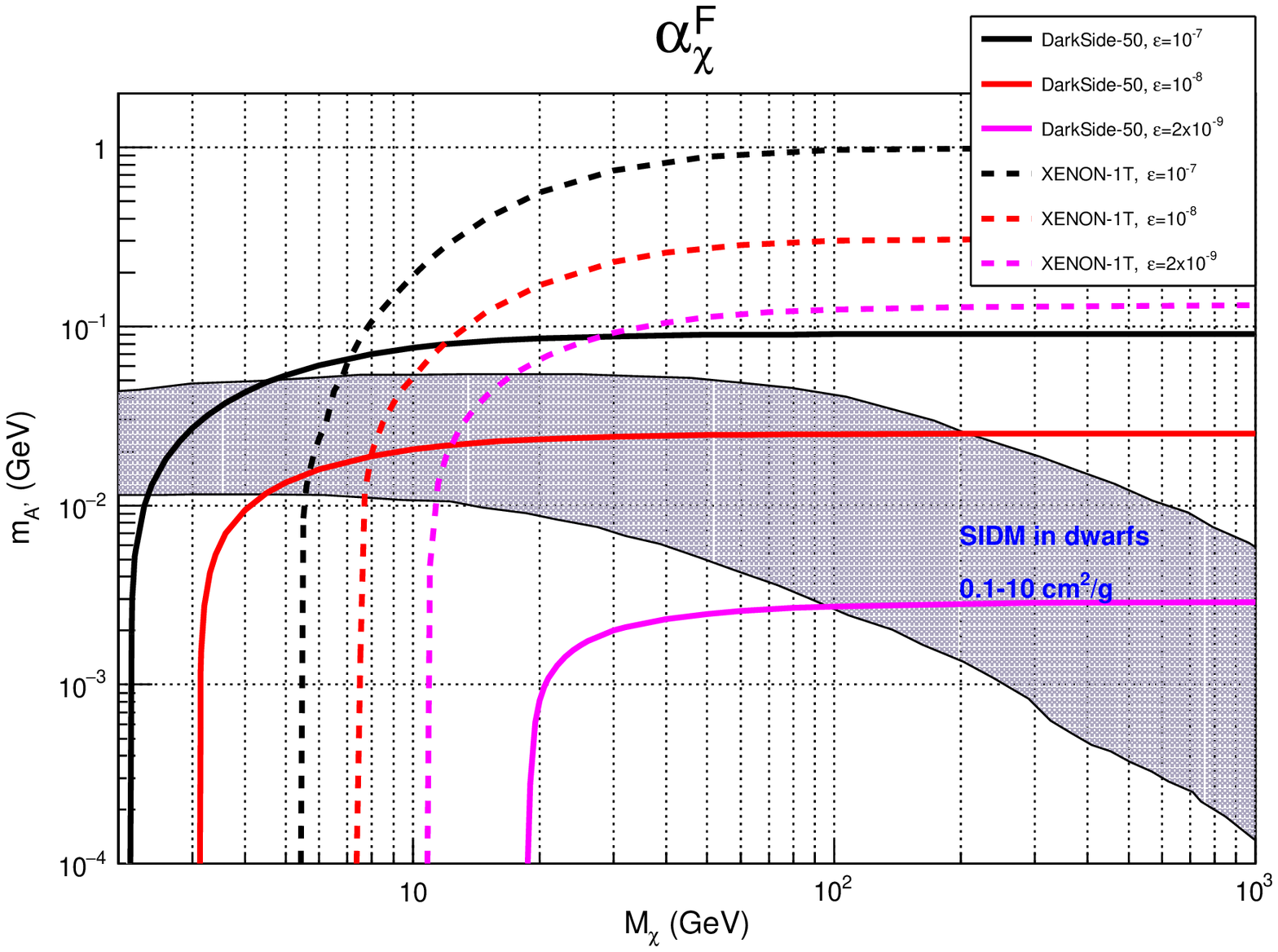}}
\hspace{0.1pt}
 \subfigure[]{
\label{2}
\includegraphics[height=8cm,width=8cm]{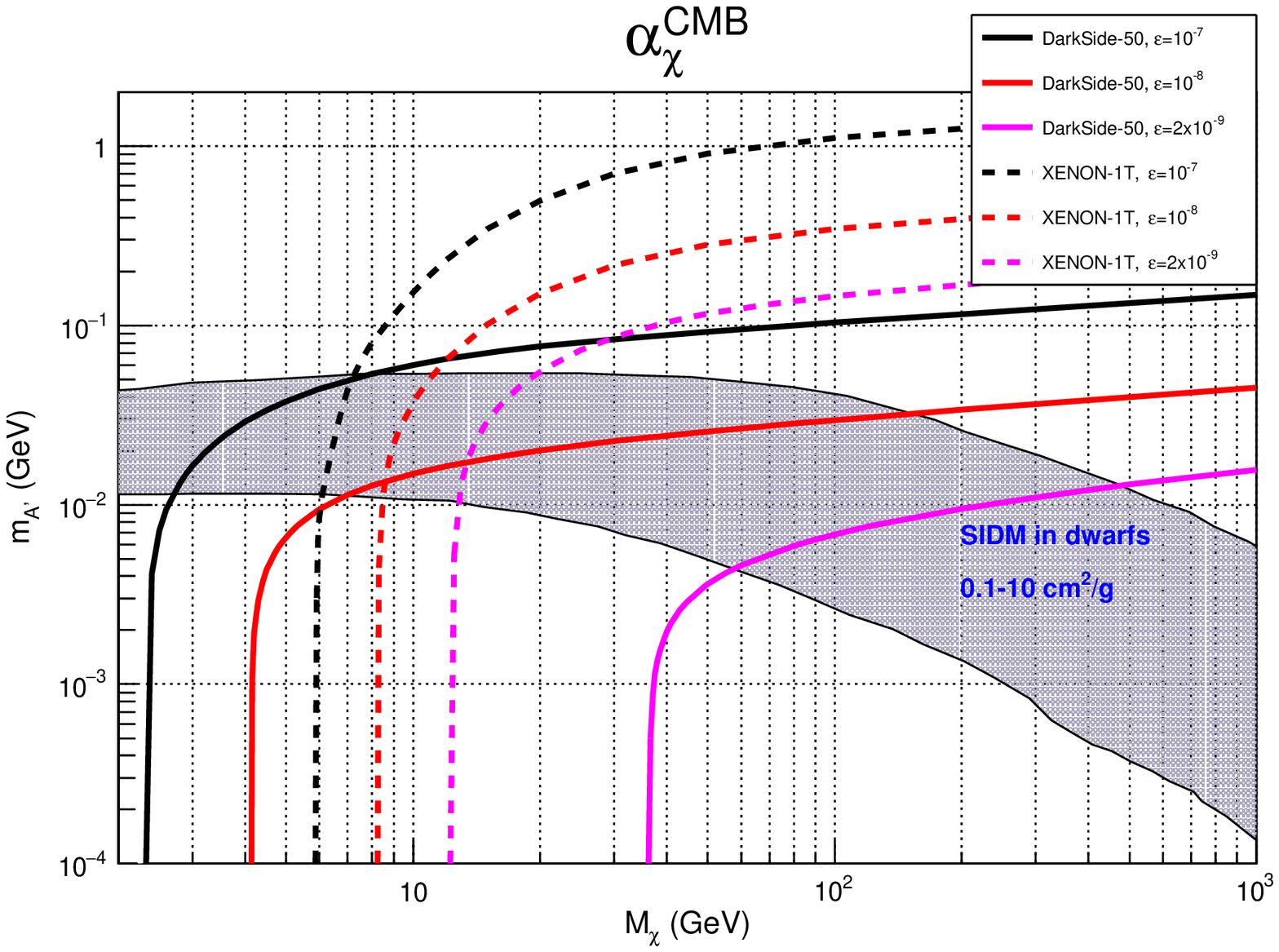}}
\caption{The XENON-1T~\cite{xenon1T:2018} and DarkSide-50~\cite{DarkSide:2018} 90\% C.L. lower limits in the ($m_{D}$, $m_{A'}$) plane, with different DM fine structure constant $\alpha_{\chi}^{{\rm F}}$(left), $\alpha_{\chi}^{{\rm CMB}}$(right). The lines are the exclusion lower limits from DarkSide-50(solid),and XENON-1T(dotted), with the different colour lines (black, red, magenta) corresponding to three $\varepsilon$ values, $10^{-7}$, $10^{-8}$, and 2$\times10^{-9}$ respectively. The same color marks the same mixing parameter, solid line (DarkSide-50) and dotted line (XENON-1T).  The shaded area is favored by observations in dwarf galaxies.}
\label{darkphoton2} 
\end{figure}

Fig.~\ref{darkphoton2} shows the direct detection constraints in the ($m_{\chi}$, $m_{A'}$) plane. The astrophysical observation gives the favored region where the self-scattering cross section per mass in dwarf galaxies is about 0.1-10 ${\rm cm^{2}/g}$. We study the constraints on DM parameters for 2 GeV $< m_{\chi} <$ 1000 GeV.
For Fig.~\ref{darkphoton2}(a), we use $\alpha_{\chi}=\alpha_{\chi}^{{\rm F}}$ to finish the analysis and find

\begin{itemize}
\item \textbf{} For $\varepsilon = 10^{-7} $, DarkSide-50 (XENON-1T) can exclude all favored region with $m_{D}\gtrsim $ 5 (7) GeV. The lower limit of exclusion from DarkSide-50 is more stringent for $m_{\chi}\lesssim$ 7 GeV.

\item\textbf{} For $\varepsilon = 10^{-8}$, DarkSide-50 (XENON-1T) can exclude all favored region with $m_{D}\gtrsim $ 200 (10) GeV. The lower limit of exclusion from DarkSide-50 is more stringent for $m_{\chi}\lesssim$ 8 GeV.

\item\textbf{} For $\varepsilon = 2\times10^{-9}$, DarkSide-50  cannot exclude the favored region with for 2 GeV $< m_{\chi} <$ 100 GeV, while the XENON-1T can exclude most of favored region  obtained by observations in dwarf galaxies for $m_{\chi} \gtrsim 20$ GeV. The lower limit of exclusion from XENON-1T is more stringent for 2 GeV $< m_{\chi} <$ 1000 GeV.

\end{itemize}
The analyses for $\alpha_{\chi}=\alpha_{\chi}^{{\rm CMB}}$ can be finished in the similar way. The results are shown in Fig.~\ref{darkphoton2}(b). Comparing (b) and (d) of Fig.~\ref{darkphoton2}, it is found that for $m_{\chi} \gtrsim$ 100 GeV, the lower limits from direct detection are more stringent with $\alpha_{\chi}=\alpha_{\chi}^{{\rm CMB}}$ than that with $\alpha_{\chi}=\alpha_{\chi}^{{\rm F}}$.

\section{Summary}

Up to now, a number of experiments have been set up to search for DM directly, and the data are accumulated. Therefore it is important to analyse these data and compare them with the theoretical predictions in order to find the existence signal of DM. In this paper, we work in an extended effective operator framework with both isospin violating interactions and light mediators, and investigate the compatibility of the candidate signal of the CDMS-II-Si with the latest constraints from DarkSide-50 and XENON-1T, etc. For the spin-independent elastic scattering, we investigate three different situations corresponding to three sets of parameters: \{$m_{\phi}$ = 200 MeV, $\xi$ = -0.7\}, \{$m_{\phi}$ = 1 MeV, $\xi$ = -0.7\}, and \{$m_{\phi}$ = 1 MeV, $\xi$ = -0.82\}, respectively. The DM mass $m_{\chi}$ favored by the CDMS-II-Si data increases when the mediator becomes lighter. The upper limits of cross section from other experiments becomes weaker and more gentle towards high DM particle mass. Fix the isospin-violation parameter $\xi $ = -0.70 (-0.82), the constraint from Xe (Ar) experiment is maximally weakened, but the favored region from CDMS-II-Si is basically excluded by XENON-1T and PandeX. We find that for isospin violating interaction with light mediator there is no parameter space which can be compatible with the positive signals from CDMS-II-Si. As a concrete example of the general DM model, we investigate the dark photon model in detail. We investigate the combined limits on the DM mass $m_{\chi}$, the dark photon mass $m_{A'}$, and the kinetic mixing parameter $\varepsilon$ in the dark photon model. In the ($m_{A'}$,$\varepsilon$) plane, we study the upper limits from several DM direct detection experiments with $m_{\chi}$ = 6 or 100 GeV. The mixing parameters $\varepsilon$ is allowed to be around $10^{-10}$ with the mediator mass range from 0.001 to 1 GeV. The upper limit obtained from different experiments have different sensitivities for various DM mass. For $m_{\chi} \lesssim 6$ GeV, the DarkSide-50 can give more stringent upper limits. For $\varepsilon = 2\times10^{-9}$, the favored region for $ m_{\chi} \lesssim $ 20 GeV is not excluded by DarkSide-50 and XENON-1T. For $\varepsilon = 10^{-8}$, DarkSide-50 (XENON-1T) can exclude all favored region with $m_{D}\gtrsim $ 200 (10) GeV, and the lower limit of exclusion from DarkSide-50 is more stringent than that from XENON-1T for $m_{\chi}\lesssim$ 8 GeV.

\section*{Acknowledgements}

This work is supported in part by the NSFC under No.11875179, No.11851303, No.11825506, No.11821505, the National Key R \& D Program of China No.2017YFA0402204, and the CAS Key research program No.XDB23030100. We thank Dr. Wei-hong Zhang for his helpful discussion. The author (Li) would like to thank the hospitality of ITP-CAS where part of this work is finished.

\end{document}